\documentclass[final,1p,times]{elsarticle}

\usepackage{multicol}
\usepackage{graphicx}
\usepackage{booktabs}
\usepackage{amssymb,bm,mathrsfs,bbm,amscd}
\usepackage[tbtags]{amsmath}
\usepackage{lastpage}

\newcommand{\be}{\begin{equation}}
\newcommand{\ee}{\end{equation}}
\newcommand{\ba}{\begin{eqnarray}}
\newcommand{\ea}{\end{eqnarray}}

\newcommand{\De}{\Delta}

\journal{Nuclear Physics A}

\begin{document}

\begin{frontmatter}



\title{Baryon Resonances}


\author{E. Oset(1), S. Sarkar(2), Bao Xi Sun(3), M. J. Vicente Vacas(1), A.
Ramos(4), P.
Gonzalez(1), J. Vijande(5), A. Martinez Torres(1), 
 K. Khemchandani(6)}

\address{1~Departamento de F\'{\i}sica Te\'orica and IFIC,
Centro Mixto Universidad de Valencia-CSIC,
Institutos de Investigaci\'on de Paterna, Aptdo. 22085, 46071 Valencia, Spain\\
2~Variable Energy Cyclotron Centre, 1/AF, Bidhannagar, Kolkata 700064, India\\
3~Institute of Theoretical Physics, College of Applied Sciences,
Beijing University of Technology, Beijing 100124, China\\
4~Departament d'Estructura i
  Constituents de la Mat\`eria and Institut de Ci\`encies del Cosmos,
  Universitat de Barcelona, 08028 Barcelona, Spain\\
5~Departamento de F\'{\i}sica Atomica Molecular y Nuclear and IFIC,
Centro Mixto Universidad de Valencia-CSIC,
Institutos de Investigaci\'on de Paterna, Aptdo. 22085, 46071 Valencia, Spain\\
6~Centro de F\'{\i}sica Computacional, Departamento de F\'{\i}sica,
Universidade de Coimbra, P-3004-516 Coimbra, Portugal\\
}

\begin{abstract}
In this talk I show recent results on how many excited baryon  resonances 
appear as systems of one meson and one baryon, or two mesons and one baryon,
with the mesons being either pseudoscalar or vectors. Connection with 
experiment is made including a discussion on old predictions and recent 
results for the photoproduction of the $\Lambda(1405)$ resonance, as well as the
prediction of one $1/2^+$ baryon state around 1920 MeV which might have been
seen in the $\gamma p \to K^+ \Lambda$ reaction.

\end{abstract}


\end{frontmatter}


\section{Introduction.}
\label{}
  The realization that the dynamics of QCD in the hadron world can be addressed
at low energies by means of effective theories in which the building blocks 
are the ground state mesons and baryons  \cite{Weinberg:1978kz} has produced
tools to address the interaction of mesons or mesons and baryons, 
mainly through chiral Lagrangians, which have had a tremendous impact in our
understanding of the spectrum of mesons and baryon resonances. We all accept that
ground states of mesons and baryons are made of $q \bar{q}$ or three $q$
respectively. Yet, the spectrum of excited hadronic states can be much richer as
we shall see.
  
  The building blocks in these chiral theories are the low energy hadrons, 
  such as the proton and baryons of its 
 SU(3) octet. To these one adds also the decuplet of the $\Delta$, considered as
 spin realignments of the three quark ground state. 
The basic mesons are the pion and mesons of its octet, to which one also adds
the nonet of the $\rho$, which also corresponds to spin realignments of the
$q \bar{q}$ ground state .

What about baryon resonances?
The logical answer is that they are excitations of the quarks, which is the
essence of quark models. This is plausible, but things could be more complicated.

Let us recall basic facts from the baryon spectrum.
The first excited $N^*$ states are the  
$N^*(1440)$ $(1/2^+)$ and the  $N^*(1535)$ $(1/2^-)$.
In quark models this will require quark excitation of around
500-600 MeV. If this is the case, one may think that it takes less energy to 
create one pion, or two (140-280 MeV). The question is whether they can be
bound or get trapped in a resonant state.
How do we know if this can occur?
We need dynamics, a potential for the interaction of mesons with
ground state baryons and then solve the Schroedinger equation
(Bethe Salpeter equation with mesons treated relativistically) 
in coupled channels. This information can be 
extracted from chiral Lagrangians: the effective
theory of QCD at low energies. This is the philosophy behind the idea of 
dynamically generated  baryons:
Many resonances are generated in this way, like the $1/2^-$ states
from meson baryon: $N^* (1535)$ \cite{Kaiser:1995eg}, two $\Lambda(1405)$ 
\cite{cola} or the $1/2^+$ states
from two mesons and a baryon, like the $N^*(1710)$ and others 
\cite{alberto1,alberto2}.

From the pseudoscalar-baryon octet interaction there are many states generated and 
one sees them as peaks in the scattering matrices or poles in the complex
plane
\cite{Kaiser:1995eg,angels,inoue,carmenjuan,hyodo,borasoyweise,borasoyulf,Oller:2006jw}.
A feature of the chiral unitary approach is its great predictive power, with the
risk that some of the predictions might not be fulfilled in Nature. 
But, so far, predictions are corroborated in production reactions, partial decay rates,
meson baryon scattering amplitudes, helicity amplitudes, 
transition form factors \cite{review,puri}.
As an example we recall in the next section some of the early calculations on photoproduction of
the $\Lambda(1405)$, taking advantage that two experiments are now reported in
this Conference ten years after the predictions were made.

\section{Photoproduction of the $\Lambda(1405)$}
  Ten years ago, with Spring8/Osaka in its initial stage, an application of the chiral
unitary approach was made to predict cross sections for photoproduction of the
$\Lambda(1405)$ \cite{Nacher:1998mi} in the $\gamma p \to K^+ \Sigma^+ \pi^-$,
$\gamma p \to K^+ \Sigma^- \pi^+$, $\gamma p \to K^+ \Sigma^0 \pi^0$ reactions. 
The cross sections found were within
measurable range, the $\Lambda(1405)$ could be clearly seen in the $\pi \Sigma$
mass distribution, as usual, and the signal was much larger than the background.
The experiment was early started at Spring8 and preliminary results were shown
in 2003 \cite{ahn}. Five years later final results were published in 
\cite{Niiyama:2008rt}, which have been reported in this Conference \cite{ahnhere}. 
Also Jeff Lab has carried out the experiment with a different set up and the results
have also been presented in this Conference \cite{moriya}.
  The model used in \cite{Nacher:1998mi} was a minimal model, in which 
only the $\gamma p K^+ P B$ vertex coming from minimal coupling, 
with P and B a pseudoscalar meson from the octet of
the $\pi$ and B a baryon of the octet of the $p$, was used. Possible effects of
baryon resonances in the $\gamma p$ entrance channels were neglected. In spite
of this, the agreement of data with the predictions is quite fair. A large
signal is seen with small background. The size of the cross section is within
50 \% of the predictions.  In adidtion, the observed $ \Sigma^+ \pi^-$,
$ \Sigma^- \pi^+$ distributions are not equal; they are shifted by a few MeV as
predicted. This is a consequence of the effect of an isospin I=1 amplitude, which
acts constructively in one case and destructively in the other. The 
$\gamma p \to K^+ \Sigma^0 \pi^0$ was predicted to be roughly the average of the
other two cross sections.  The details of the recent experiments can be seen in
the devoted talks  \cite{ahnhere,moriya} and are rather interesting. A
remarkable thing is that the peak positions for $ \Sigma^+ \pi^-$,
$ \Sigma^- \pi^+$ production seem to be reversed in \cite{moriya} than
predicted,
and in \cite{ahnhere} they seem to be angle dependent, with opposite trends at
forward and backward angles.  

   The new  experimental information obtained calls for a theoretical revival of
the theory to the light of the findings made in chiral unitary approaches in the
last decade. At stake are issues like the nature of the $\Lambda(1405)$
resonance, the existence of the two $\Lambda(1405)$ resonances for which
 experimental evidence has been claimed \cite{magas,jidoseki}, and the
possibility that the I=1 amplitude, which is clearly visible in the different 
$ \Sigma^+ \pi^-$, $ \Sigma^- \pi^+$ cross sections, could be of resonant
character, evidencing a new $\Sigma$ resonance around 1400 MeV, for which hints
were seen in \cite{cola,ollerulf}. Recent claims for this resonance have been
made in  \cite{wuzou} from the study of the $K^-p \to \Lambda \pi^+ \pi^-$
reaction.

\section{Resonances from the interaction of vector mesons with baryons}

  This is a very novel development since, as we shall see, some of the high mass
baryon resonances can be represented like bound states of vector mesons and
baryons, either from the octet of stable baryons or the decuplet.

\subsection{Formalism}

We follow the formalism of the hidden gauge interaction for vector mesons of
\cite{hidden1,hidden2,hidden3} (see also \cite{hidekoroca} for a practical set of Feynman rules).
The  Lagrangian involving the interaction of
vector mesons amongst themselves is given by
\begin{equation}
{\cal L}_{III}=-\frac{1}{4}\langle V_{\mu \nu}V^{\mu\nu}\rangle \ ,
\label{lVV}
\end{equation}
where the symbol $\langle \rangle$ stands for the trace in the $SU(3)$ space
and $V_{\mu\nu}$ is given by
\begin{equation}
V_{\mu\nu}=\partial_{\mu} V_\nu -\partial_\nu V_\mu -ig[V_\mu,V_\nu]\ ,
\label{Vmunu}
\end{equation}
where  $g$ is  $g=\frac{M_V}{2f}$,
with $f=93$ MeV the pion decay constant.  The magnitude $V_\mu$ is the ordinary 
$SU(3)$ matrix of the vectors of the octet of the $\rho$

The lagrangian ${\cal L}_{III}$ gives rise to a contact term coming from
$[V_\mu,V_\nu][V_\mu,V_\nu]$, as well as to a three
vector vertex 
\begin{equation}
{\cal L}^{(c)}_{III}=\frac{g^2}{2}\langle V_\mu V_\nu V^\mu V^\nu-V_\nu V_\mu
V^\mu V^\nu\rangle\ ;~~{\cal L}^{(3V)}_{III}
=ig\langle (V^\mu\partial_\nu V_\mu -\partial_\nu V_\mu
V^\mu) V^\nu\rangle,
\label{lcont}
\end{equation}

In this case one finds an analogy to the coupling of vectors to
 pseudoscalars given in the same theory by
\begin{equation}
{\cal L}_{VPP}= -ig \langle [
P,\partial_{\nu}P]V^{\nu}\rangle \ ,
\label{lagrVpp}
\end{equation}
where $P$ is the SU(3) matrix of the pseudoscalar fields.

In a similar way, one obtains the Lagrangian for the coupling of vector mesons to
the baryon octet given by
\cite{Klingl:1997kf,Palomar:2002hk} \footnote{Correcting a misprint in
\cite{Klingl:1997kf}}
\begin{equation}
{\cal L}_{BBV} =
g\left( \langle \bar{B}\gamma_{\mu}[V^{\mu},B]\rangle +
\langle \bar{B}\gamma_{\mu}B \rangle \langle V^{\mu}\rangle \right)
\label{lagr82}
\end{equation}
where $B$ is now the ordinary SU(3) matrix of the baryon octet

\begin{figure}[ht!]
\begin{center}
\includegraphics[width=0.4\textwidth]{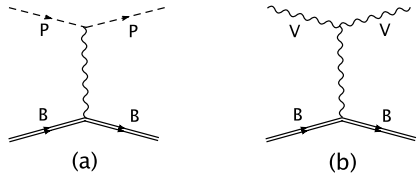}
\caption{Diagrams contributing to the pseudoscalar-baryon (a) or vector-
baryon (b) interaction via the exchange of a vector meson.}
\label{fig:feyn}
\end{center}
\end{figure}

With these ingredients we can construct the Feynman diagrams that lead to the $PB
\to PB$ and $VB \to VB$ interaction, by exchanging a vector meson between the
pseudoscalar or the vector meson and the baryon, as depicted in Fig.
\ref{fig:feyn}.

From the diagram of Fig. \ref{fig:feyn}(a), and under the low energy approximation of
neglecting $q^2/M_V^2$ in the propagator of the exchanged vector, where $q$ is the
momentum transfer, one obtains the
same amplitudes as obtained from the ordinary chiral Lagrangian for
pseudoscalar-baryon octet interaction, namely the Weinberg-Tomozawa
terms.  The approximation of neglecting the three momenta of the vectors implies
that $V^{\nu}$ in eq. (\ref{lcont}) corresponds to the exchanged vector and the analogy
with eq. (\ref{lagrVpp}) is more apparent. Note that $\epsilon_\mu \epsilon^\mu$ becomes
$-\vec{\epsilon}\,\vec{\epsilon }\,^\prime$ and the signs of the Lagrangians also agree.

   A small amendment is in order in the case of vector mesons, which
   is due to the mixing of $\omega_8$ and the singlet of SU(3), $\omega_1$, to give the
   physical states of the $\omega$ and the $\phi$ mesons. The practical rule is
   simple and can be found in \cite{angelsvec}.  Upon the approximation consistent with
   neglecting the three momentum versus the mass of the particles (in this
   case the baryon), we can just take the $\gamma^0$ component of
   Eq. (\ref{lagr82})  and
   then the transition potential corresponding to the diagram of \ref{fig:feyn}(b) is
   given by
   \begin{equation}
V_{i j}= - C_{i j} \, \frac{1}{4 f^2} \, \left( k^0 + k^\prime{}^0\right)
~\vec{\epsilon}\,\vec{\epsilon }\,^\prime
\label{kernel}
\end{equation}
   where $k^0, k^\prime{}^0$ are the energies of the incoming and outgoing vector meson.
The $C_{ij}$ coefficients of eq. (\ref{kernel}) can also be found in
    \cite{angelsvec}, where 
     one can see that the
     cases with $(I,S)=(3/2,0)$, $(2,-1)$ and $(3/2,-2)$, the last two
     corresponding to exotic channels, have a repulsive interaction
and do not produce poles in the scattering matrices.  However, the sectors
$(I,S)=(1/2,0)$, $(0,-1)$, $(1,-1)$ and $(1/2,-2)$ are attractive and one finds
 bound states and resonances in these cases.

    The scattering matrix is obtained  solving the
    coupled channels Bethe Salpeter equation in the on shell factorization approach of
    \cite{angels,ollerulf}
   \begin{equation}
T = [1 - V \, G]^{-1}\, V
\label{eq:Bethe}
\end{equation}
with $G$ being the loop function of a vector meson and a baryon. This function
is convoluted with the spectral function of the vector mesons to take into
account their width as done in \cite{nagahiro}.

 In this
case the factor $\vec{\epsilon}\,\vec{\epsilon }\,^\prime$, appearing in the potential $V$,
factorizes also in the $T$ matrix for the external vector mesons. This trivial
spin structure is responsible for having degenerate states with spin-parity
$1/2^-, 3/2^-$ for the interaction of vectors with the octet of baryons and
$1/2^-, 3/2^-, 5/2^-$ for the interaction of vectors with the decuplet
 of baryons.

  What we have done here for the interaction of vectors with the octet of
  baryons can be done for the interaction of vectors with the decuplet of
  baryons, and the interaction is obtained directly from that of the
  pseudoscalar-decuplet of baryons studied in
  \cite{kolodecu,Sarkar:2004jh}. The study of this interaction
  in \cite{vijande,souravbao} leads
  also to the generation of many resonances which are described below.

We search for poles in the scattering matrices in the second Riemann sheet, as
defined in previous works \cite{luisaxial}.

\subsection{Results}

In table \ref{tab:pdg} we show a summary of the results obtained from the
interaction of vectors with the octet of baryons \cite{angelsvec} and the tentative
association to known states \cite{pdg}.

\begin{table}[ht!]
\begin{center}
      \renewcommand{\arraystretch}{1.5}
     \setlength{\tabcolsep}{0.2cm}
\begin{tabular}{c|c|cc|ccccc}\hline\hline
$I,\,S$&\multicolumn{3}{c|}{Theory} & \multicolumn{5}{c}{PDG data}\\
\hline
    \vspace*{-0.3cm}
    & pole position    & \multicolumn{2}{c|}{real axis} &  &  & &  &  \\
    &   & mass & width &name & $J^P$ & status & mass & width \\
    \hline
$1/2,0$ & --- & 1696  & 92  & $N(1650)$ & $1/2^-$ & $\star\star\star\star$ & 1645-1670
& 145-185\\
  &      &       &     & $N(1700)$ & $3/2^-$ & $\star\star\star$ &
	1650-1750 & 50-150\\
       & $1977 + {\rm i} 53$  & 1972  & 64  & $N(2080)$ & $3/2^-$ & $\star\star$ & $\approx 2080$
& 180-450 \\	
   &     &       &     & $N(2090)$ & $1/2^-$ & $\star$ &
 $\approx 2090$ & 100-400 \\
 \hline
$0,-1$ & $1784 + {\rm i} 4$ & 1783  & 9  & $\Lambda(1690)$ & $3/2^-$ & $\star\star\star\star$ &
1685-1695 & 50-70 \\
  &       &       &    & $\Lambda(1800)$ & $1/2^-$ & $\star\star\star$ &
1720-1850 & 200-400 \\
       & $1907 + {\rm i} 70$ & 1900  & 54  & $\Lambda(2000)$ & $?^?$ & $\star$ & $\approx 2000$
& 73-240\\
       & $2158 + {\rm i} 13$ & 2158  & 23  &  &  &  & & \\
       \hline
$1,-1$ &  ---  & 1830  & 42  & $\Sigma(1750)$ & $1/2^-$ & $\star\star\star$ &
1730-1800 & 60-160 \\
  & ---    & 1987  & 240  & $\Sigma(1940)$ & $3/2^-$ & $\star\star\star$ & 1900-1950
& 150-300\\
   &     &       &   & $\Sigma(2000)$ & $1/2^-$ & $\star$ &
$\approx 2000$ & 100-450 \\\hline
$1/2,-2$ & $2039 + {\rm i} 67$ & 2039  & 64  & $\Xi(1950)$ & $?^?$ & $\star\star\star$ &
$1950\pm15$ & $60\pm 20$ \\
         & $2083 + {\rm i} 31 $ &  2077     & 29  &  $\Xi(2120)$ & $?^?$ & $\star$ &
$\approx 2120$ & 25  \\
 \hline\hline
    \end{tabular}
\caption{The properties of the 9 dynamically generated resonances and their possible PDG
counterparts.}
\label{tab:pdg}
\end{center}
\end{table}

  For the $(I,S)=(1/2,0)$ $N^*$  states there is the $N^*(1700)$ with
 $J^P=3/2^-$, which could correspond to the state we find with the same quantum
 numbers around the same energy. We also find in the PDG the  $N^*(1650)$, which
 could be the near degenerate spin parter of the $N^*(1700)$ that we predict in
 the theory. It is interesting to recall that in the study of
 Ref.~\cite{mishajuelich} a pole is found around 1700 MeV,
with the largest coupling to $\rho N$ states.
Around 2000 MeV, where we find another $N^*$ resonance,
there are the states $N^*(2080)$ and $N^*(2090)$, with $J^P=3/2^-$ and
$J^P=1/2^-$ respectively, showing a good approximate spin degeneracy.

For the case $(I,S)=(0,-1)$ there is in the PDG one state, the $\Lambda(1800)$
with $J^P=1/2^-$, remarkably close to the energy were we find a $\Lambda$
state.  The state obtained around 1900 MeV could
correspond to the $\Lambda(2000)$ cataloged in the PDG with unknown spin and parity.

 The case of the $\Sigma $ states having $(I,S)=(1,-1)$ is rather interesting.
 The state
that we find around 1830 MeV, could be associated to the  $\Sigma(1750)$
with $J^P=1/2^-$. More interesting seems to be the case of the state obtained around
1990 MeV that could be related to two PDG candidates, again
nearly degenerate, the $\Sigma(1940)$ and the $\Sigma(2000)$, with spin and
parity  $J^P=3/2^-$ and $J^P=1/2^-$ respectively.

  Finally, for the case of the cascade resonances, $(I,S)=(1/2,-2)$, we find
  two states, one  around 2040 MeV and the other one around 2080 MeV. There are two cascade states in
  the PDG around this energy region with spin parity unknown, the
  $\Xi(1950)$ and the $\Xi(2120)$.  Although the experimental
  knowledge of this sector is relatively poor, a program is presently running at
  Jefferson Lab to improve on this situation \cite{Nefkens:2006bc}.

    The case of the vector interaction with the decuplet is
similar \cite{Sarkar:2004jh} and we show the results in Table \ref{tab:pdg2}

\begin{table*}[!ht]
      \renewcommand{\arraystretch}{1.5}
     \setlength{\tabcolsep}{0.2cm}
\begin{center}
\begin{tabular}{c|l|cc|lcclc}\hline\hline
$S,\,I$&\multicolumn{3}{c|}{Theory} & \multicolumn{5}{c}{PDG data}\\\hline
        & pole position &\multicolumn{2}{c|}{real axis} & name & $J^P$ & status & mass & width \\
        &               & mass & width & \\\hline
$0,1/2$ & $1850+i5$   & 1850  & 11  & $N(2090)$ & $1/2^-$ & $\star$ & 1880-2180 & 95-414\\
        &             &       &     & $N(2080)$ & $3/2^-$ & $\star\star$ & 1804-2081 & 180-450\\
        &       &  $2270(bump)$ &  & $N(2200)$ & $5/2^-$ & $\star\star$ & 1900-2228 & 130-400\\
\hline
$0,3/2$ & $1972+i49$  & 1971  & 52  & $\De(1900)$ & $1/2^-$ & $\star\star$ & 1850-1950 & 140-240 \\
    &             &       &     & $\De(1940)$ & $3/2^-$ & $\star$ & 1940-2057 & 198-460   \\
        &             &       &     & $\De(1930)$ & $5/2^-$ & $\star\star\star$ & 1900-2020  & 220-500   \\
    &             & $2200 (bump)$  &     & $\De(2150)$ & $1/2^-$ & $\star$ & 2050-2200  & 120-200  \\
\hline
$-1,0$  & $2052+i10$  & 2050  & 19  & $\Lambda(2000)$ & $?^?$ & $\star$  & 1935-2030 & 73-180\\
\hline
$-1,1$  & $1987+i1$   & 1985  & 10   & $\Sigma(1940)$ & $3/2^-$  & $\star\star\star$ &
1900-1950 & 150-300 \\
        & $2145+i58$  & 2144  & 57  & $\Sigma(2000)$ & $1/2^-$  & $\star$ & 1944-2004 &
    116-413\\
    & $2383+i73$  & 2370  & 99 & $\Sigma(2250)$ & $?^?$ & $\star\star\star$ & 2210-2280 &
    60-150\\
    &   &   &  & $\Sigma(2455)$ & $?^?$ & $\star\star$ & 2455$\pm$10 &
    100-140\\
\hline
$-2,1/2$ & $2214+i4$  & 2215  & 9  & $\Xi(2250)$ & $?^?$ & $\star\star$ & 2189-2295 & 30-130\\
     & $2305+i66$ & 2308  & 66 & $\Xi(2370)$ & $?^?$ & $\star\star$ & 2356-2392 & 75-80 \\
         & $2522+i38$ & 2512  & 60  & $\Xi(2500)$ & $?^?$ & $\star$ & 2430-2505 & 59-150\\
\hline
$-3,0$   & $2449+i7$   & 2445 & 13  & $\Omega(2470)$   & $?^?$   & $\star\star$ & 2474$\pm$12 & 72$\pm$33\\
 \hline\hline
    \end{tabular}
\caption{The properties of the 10 dynamically generated resonances and their possible PDG
counterparts. We also include the $N^*$ bump around 2270 MeV and the $\Delta^*$ bump around 2200 MeV. }
\label{tab:pdg2}
\end{center}
\end{table*}

 We also can see that in many cases the experiment shows the near degeneracy
 predicted by the theory. Particularly, the case of the three $\Delta$
 resonances around 1920 MeV is very interesting. One observes a near
 degeneracy in the three spins $1/2^-, 3/2^-, 5/2^-$, as the theory predicts. It
 is also very instructive to recall that the case of the  $\Delta(5/2^-)$ is
 highly problematic in quark models since it has a  $3~h\omega$ excitation
 and comes out always with a very high mass \cite{vijande,pedro}.

  The association of states found to some resonances reported in the PDG
  for the case of $\Lambda$, $\Sigma $ and $\Xi$ states looks also equally
  appealing as one can see from the table.

  The reasonable results reported here produced by the hidden gauge approach
   should give a stimulus to
  search experimentally for the missing spin partners of the already observed
  states, as well as possible new ones.

\section{States of two mesons and a baryon}
 There are two specific talks on this issue in the Workshop
 \cite{albertotalk,jidotalk}. I
 will summarize a bit the important findings in this area by different groups.
In \cite{alberto1,alberto2} a formalism was developed to study Faddeev equations
of systems of two mesons and a stable baryon. The interaction of the pairs
was obtained from the chiral unitary approach, which proves quite successful to
give the scattering amplitudes of meson-meson and meson-baryon systems in the
region of energies of interest to us. The spectacular finding is that,
leaving apart the Roper resonance, whose structure is far more elaborate than
originally thought \cite{Krehl:1999km,Dillig:2004rh}, all the low lying
$J^P=1/2^+$ excited states are obtained as bound states or resonances of two
mesons and one baryon in coupled channels.

\begin{figure}[ht!]
\begin{center}
\includegraphics[scale=0.7]{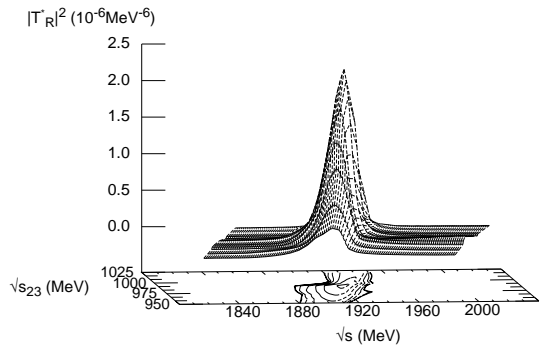}
\caption{A possible $N^*(1910)$ in the $N K\bar{K}$ channels.}
\label{threebody}
\end{center}
\end{figure}

  Particularly relevant to this Conference is the issue of a possible bound 
state of $K \bar{K} N$. In \cite{jidothree},
using variational methods, the authors found a bound state of $K \bar{K} N$, with
the $K \bar{K}$ being in the $a_0(980)$ state \cite{jidothree}.  The system
was studied a posteriori in \cite{albertopheno} and it was found to
appear at the same energy and the same configuration, although with a mixture
of $f_0(980) N$, see fig. \ref{threebody}. This state appears around 1920 MeV with $J^P=1/2^+$. In a
recent paper \cite{albertoulf} some arguments were given to associate this state
with the bump that one sees in the $\gamma p \to K^+ \Lambda$ reaction around
this energy, which is clearly visible in recent accurate experiments
\cite{Bradford:2005pt,Sumihama:2005er}. If this association was correct there
would be other  experimental consequences, as an enhanced strength of the
$\gamma p \to K^+ K^- p$ cross section close to threshold, as well as a shift
of strength close to the $K \bar{K}$ threshold in the invariant mass
distribution of the kaon pair.  This experiment is right now under study 
\cite{nakanotalk}. Another
suggestion of \cite{albertoulf} is to measure the total $\gamma p$ spin 
$S_z=1/2$ and $S_z=3/2$
amplitudes, the $z$-direction along the photon momentum, since this 
would discriminate the cases where the peak around 
1920 MeV is due to a $1/2^+$ or a $3/2^+$ resonance.

\section{Acknowledgments}
This work is partly supported by the EU contract No. MRTN-CT-2006-035482
(FLAVIAnet), by the contracts FIS2006-03438 FIS2008-01661 from MICINN
(Spain) and by the Ge\-ne\-ra\-li\-tat de Catalunya contract 2005SGR-00343. We
acknowledge the support of the European Community-Research Infrastructure
Integrating Activity ``Study of Strongly Interacting Matter'' (HadronPhysics2,
Grant Agreement n. 227431) under the Seventh Framework Programme of EU.



\end{document}